\pdfoutput=1
\documentclass[sigconf, nonacm]{acmart}

\AtBeginDocument{%
  }

\copyrightyear{2024}
\acmYear{2024}
\setcopyright{rightsretained}
\acmConference[DAC '24]{61st ACM/IEEE Design Automation Conference}{June 23--27, 2024}{San Francisco, CA, USA}
\acmBooktitle{61st ACM/IEEE Design Automation Conference (DAC '24), June 23--27, 2024, San Francisco, CA, USA}
\acmDOI{10.1145/3649329.3657341}
\acmISBN{979-8-4007-0601-1/24/06}

\usepackage{subcaption}
\begin{document}

\title{Efficient Open Modification Spectral Library Searching in High-Dimensional Space with Multi-Level-Cell Memory}

\settopmatter{authorsperrow=1}
\newcommand{\tsc}[1]{\textsuperscript{#1}} 
\author{Keming Fan\tsc{1*}, Wei-Chen Chen\tsc{2}, Sumukh Pinge\tsc{1}, 
H.-S. Philip Wong\tsc{2}, Tajana Rosing\tsc{1*}}
\affiliation{
  \institution{\tsc{1}University of California, San Diego, \tsc{2}Stanford University}
  \country{}
}
\email{*{k4fan, tajana}@ucsd.edu}

\renewcommand{\shortauthors}{Fan et al.}

\begin{abstract}
Open Modification Search (OMS) is a promising algorithm for mass spectrometry analysis that enables the discovery of modified peptides. However, OMS encounters challenges as it exponentially extends the search scope. Existing OMS accelerators either have limited parallelism or struggle to scale effectively with growing data volumes. In this work, we introduce an OMS accelerator utilizing multi-level-cell (MLC) RRAM memory to enhance storage capacity by 3x. Through in-memory computing, we achieve up to 77x faster data processing with two to three orders of magnitude better energy efficiency. Testing was done on a fabricated MLC RRAM chip. We leverage hyperdimensional computing to tolerate up to 10\% memory errors while delivering massive parallelism in hardware.

\end{abstract}



\maketitle

\section{Introduction}

Mass spectrometry (MS) analysis is a critical technique for studying proteins, which serve as the fundamental components of modern medicines. In a typical MS experiment, numerous spectra, known as query spectra, are generated and compared with a reference database containing known peptides. Proteins in MS experiments often undergo post-translational modifications (PTM), altering mass and properties. However, the reference database only includes spectra for unmodified peptides. This complicates the search, as many peptides may not find a match.

Open modification search (OMS) offers a promising solution to circumvent this issue by allowing the identification of modified spectra.  
In the traditional standard search, comparisons are limited to query spectra and reference spectra that share similar precursor mass.
In contrast, OMS extends the matching scope to a broader range. It adopts a wide precursor mass window on reference spectra, which accounts for the mass shifts induced by PTMs and other protein modifications.
This approach enables the comparison of spectra from modified proteins with their unmodified counterparts, thereby facilitating a more comprehensive analysis.

However, OMS encounters challenges as it vastly expands the search scope, necessitating a more refined design. Specifically, we require (1) dense memory solutions given the exponentially growing size of data, and (2) algorithms that are easily parallelizable in hardware for faster data processing. Several works have accelerated the OMS algorithm.
The ANN-SoLo tool  \cite{ANN_SoLo'23} uses nearest neighbor indexing to select candidates and employs the shifted dot product to compute scores on those candidates. Nevertheless, ANN-SoLo demonstrates limited data parallelism as it uses complicated high-precision floating-point arithmetic. 
HyperOMS  \cite{HyperOMS} encodeds input data into high dimensional vectors and performs simple integer operations. This approach results in significant increase in parallelism on GPU architectures. However, both of them do not scale well with the growing data volumes, necessitating high-density memory solutions. In this work, we use dense multi-level-cell (MLC) RRAM to increase the storage capacity.
In addition, we employ an in-memory computing approach to reduce data movement, leading to faster data processing. Since MLC RRAM and in-memory computing are usually error prone, we leverage the robustness of hyperdimensional computing (HD) to tolerate these errors.
The main contributions of this work are as follows.
\begin{itemize}
    \item An OMS accelerator using HD and MLC RRAM is proposed. The proposed design achieves 3x better storage capacity per area with comparable accuracy to state-of-the-art, allowing for up to 10\% memory error tolerance.
    \item We accelerate the main stages of the algorithm by processing in memory. The functionality is tested through experiments on a fabricated MLC RRAM chip.
    \item We propose several hardware-software co-design strategies, including a multi-bit hypervector scheme and an efficient mapping scheme to enhance computational efficiency. 
    
\end{itemize}

\section{Related Work and Motivation}
\subsection{Open Modification Search}

Mass Spectrometry (MS) is crucial in proteomic research, enabling the analysis of complex biological samples. 
Open Modification Search (OMS) marks a significant evolution in MS technology, as it allows for the discovery of modified peptides.
However, the implementation of OMS is challenging as it significantly enlarges search space. This expanded scope includes both unmodified and modified peptide variants, leading to increased computational demands. HyperOMS, the fastest existing OMS accelerator that operates on GPUs, still faces a challenge with a large memory footprint \cite{HyperOMS}. This challenge arises from OMS's inherent memory-intensive nature, leading to efficiency concerns and data transfer bottlenecks. Processing in memory enables direct computations within the memory space, offering a better solution for OMS acceleration.

\subsection{RRAM and In-memory Computing}

Another challenge arises from the escalating data volumes. The public data for mass spectrometry analysis is experiencing exponential growth \cite{perez2022pride}; however, existing memory solutions face difficulties in scaling to meet the expanding demands. 

Resistive random access memory (RRAM) is an emerging non-volatile memory that stores data by changing its resistance. 
In TSMC 22nm technology, a single-level-cell (SLC) RRAM provides 3x higher storage capacity per area than high-density SRAM (Static Random Access Memory) \cite{chou202022nm}, positioning it as an optimal solution for extensive data storage requirements.

\begin{figure}[h]
    \vspace{-10pt}
    \centering
    \begin{subfigure}[t]{0.55\columnwidth}
        \centering
        \includegraphics[width=\textwidth]{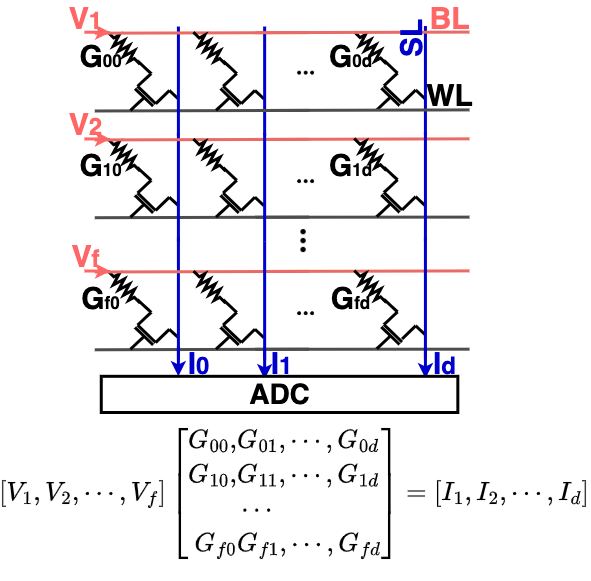}
        \subcaption{Crossbar array for MVM}
        \label{rram_pim}
    \end{subfigure}
    \hfill
    \begin{subfigure}[t]{0.44\columnwidth}
        \centering
        \includegraphics[width=\textwidth]{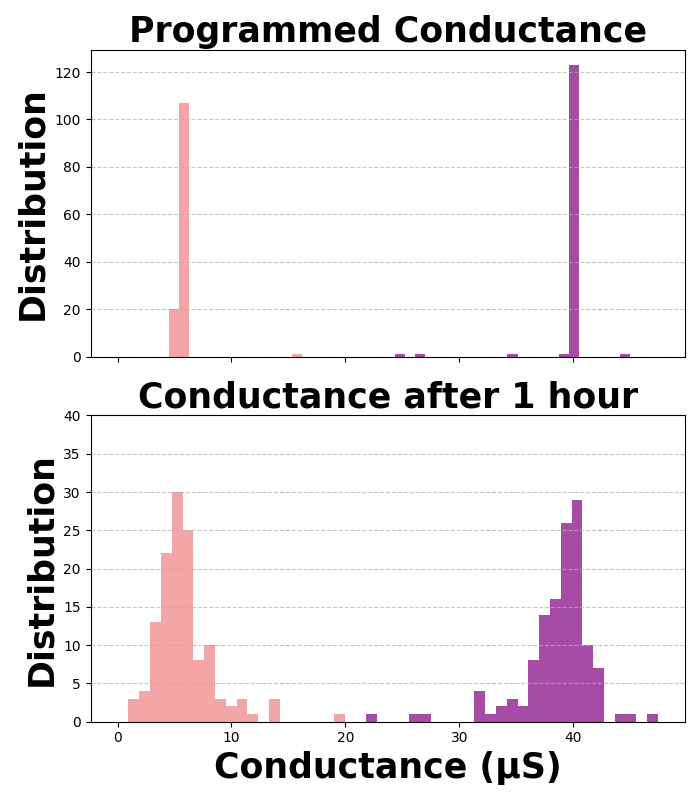}
        \subcaption{Conductance relaxation}
        \label{conductance_relax}
    \end{subfigure}
    \caption{RRAM}
    \vspace{-6pt}
\end{figure}

Despite its density advantage, RRAM can effectively contribute to parallel in-memory acceleration.
Prior research has explored the computational power of RRAM, particularly for the MAC (multiply-accumulate) operation that is critical in modern neural networks. Figure \ref{rram_pim} shows a typical configuration \cite{wan2022compute} of 1T1R RRAM array designed for matrix vector multiplication (MVM). Input data is mapped to analog voltages, and RRAM-stored weights are represented by their conductance. Currents are generated across multiple rows and accumulated along the columns, following Kirchhoff's law. The ADC (Analog-to-Digital Converter) then converts currents to digital values, representing the computation outputs.

Previous works  \cite{xu2023fsl, xue201924, chen201865nm} have primarily utilized SLC RRAM for storage or computation in various applications. In SLC memory, each RRAM device is programmable to low resistance to represent '1' or to high resistance to represent '0'. However, the potential of RRAM extends beyond binary storage, as it can achieve arbitrary analog resistance states by applying different voltages, enabling the storage of multi-bit data. This configuration is commonly referred to as multi-level-cell (MLC) RRAM.
However, adopting MLC RRAM, while promising in increasing storage capacity, poses challenges due to device non-idealities. RRAM suffers from conductance relaxation and a relatively low on-off ratio. Figure \ref{conductance_relax} illustrates the conductance distribution in RRAM collected from a fabricated chip 
  \cite{wan2022compute} after 60 minutes of programming, displaying a shift in conductance that hinders accurate storage and computing. Some works attempt to harness the potential of MLC RRAM.  The authors of 
  \cite{carsello2022amber} use RRAM as a passive array for storage only, without the computing ability. In 
  \cite{li202240}, they proposed an RRAM-based in-memory computing macro, but with only three levels per cell, which doesn't fully use the potential of MLC RRAM. 
Motivated by these challenges and recognizing the limitations of existing approaches, we propose a robust algorithm using hyperdimensional computing to tolerate errors associated with MLC RRAM.

\section{Algorithm Overview}

Hyperdimensional computing (HD) is a brain-inspired computing method that emulates neuronal activity. HD encodes information to binary high-dimensional (long) vectors called hypervectors, typically with a dimension of 1k-10k \cite{kanerva2009hyperdimensional}. In this work, we leverage HD for OMS acceleration, benefiting from its high degree of parallelism in hardware implementations and robustness to errors.

\begin{figure}[h]
    \vspace{-4pt}
    \centering
    \includegraphics[width=.49\textwidth]{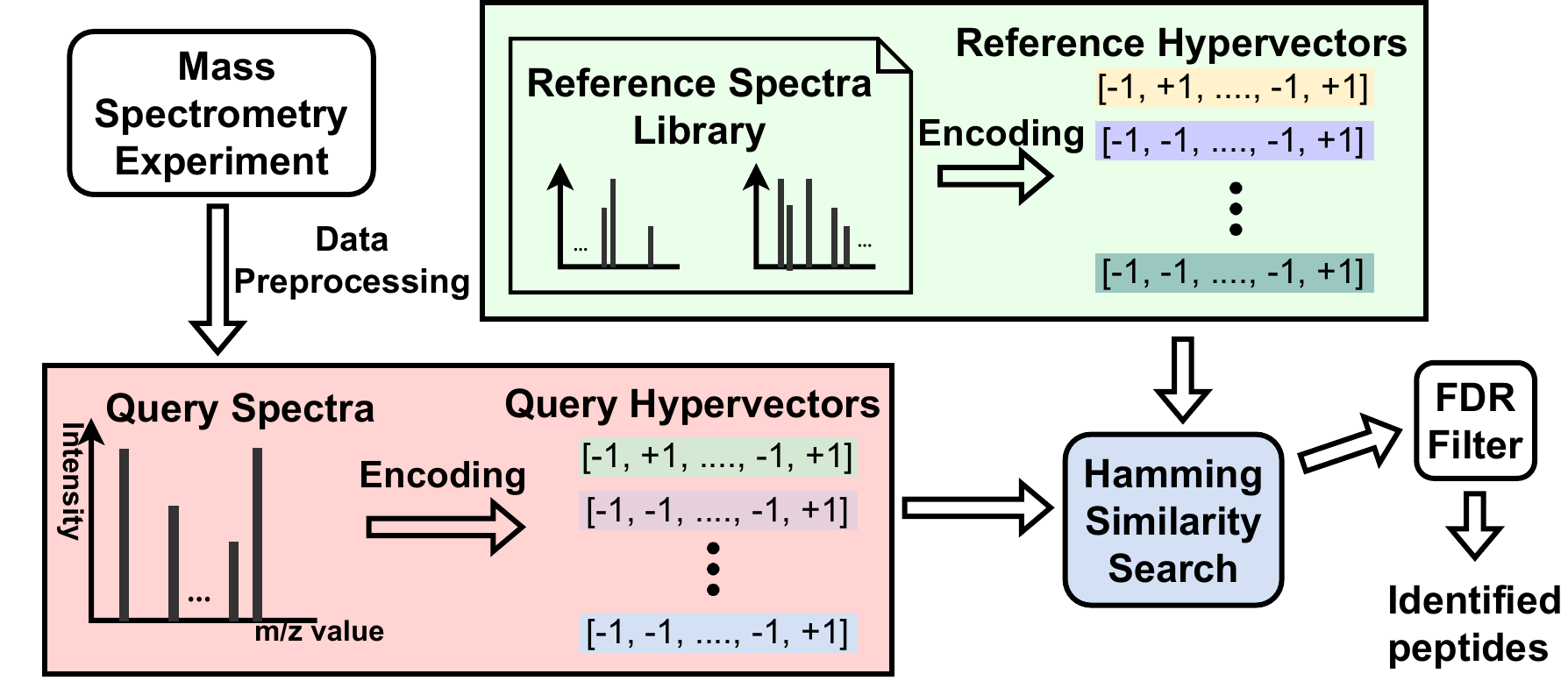}
     \caption{Overall flow}
     \label{overview}
    \vspace{-8pt}
\end{figure}

Figure \ref{overview} illustrates the overall diagram, which includes data preprocessing that turns raw data into spectra vectors. This is followed by HD encoding and hamming search in high-dimensional space (hyperspace). Finally, an FDR filter outputs identified peptides.

\subsection{Data Preprocessing} Preprocessing is the initial step in MS analysis, beginning with the extraction of prominent peak features from raw data. This involves identifying and retaining peaks above a predefined intensity threshold, typically set at 1\% of the greatest peak intensity. The goal is to eliminate background noise, resulting in a refined set of 50 to 150 peaks representing the most useful features in each spectrum.
Next, spectra are transformed into vectors by categorizing mass-to-charge (m/z) ratios into bins. The resulting vectors contain floating-point values reflecting peak intensities. In cases where multiple peaks fall within a bin, their intensities are summed.

\subsection{Encoding}
HD encodes spectra vectors into orthogonal binary hypervectors that represent unique features within the spectra. Previous research explored various encoding methods, such as permutation-based \cite{salamat2019f5} and random projection encoding \cite{cannings2017random}. However, these methods may not effectively capture key features, such as m/z values and peak intensities in the spectra. In this work, we employ an ID-Level encoding method \cite{imani2017voicehd} to address this limitation.

\begin{figure}[h]
    \vspace{-8pt}
    \centering
    \includegraphics[width=0.28\textwidth]{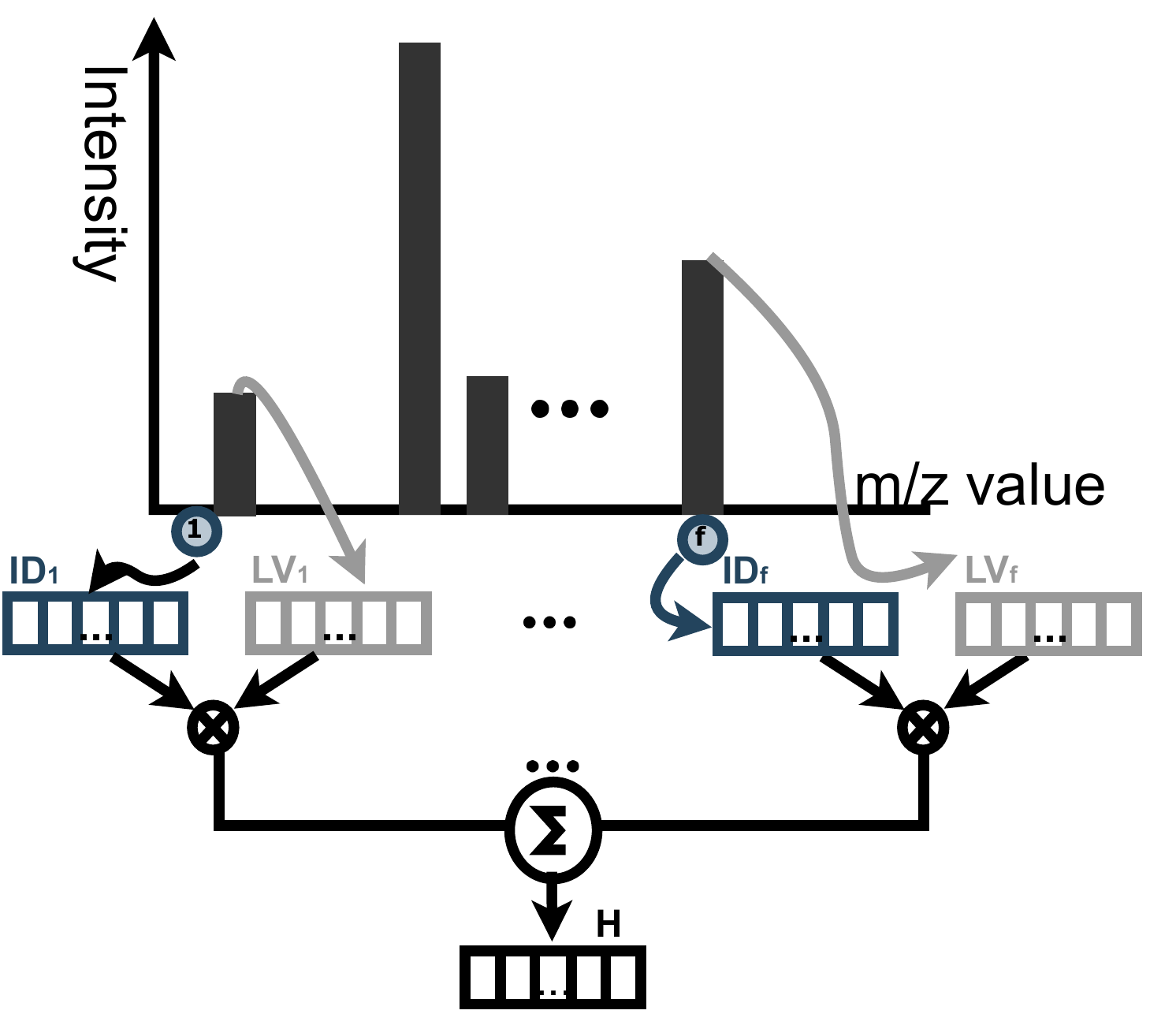}
    \caption{ID-Level Encoding}
    \label{encode}
    \vspace{-10pt}
\end{figure}

During encoding, each peak position (m/z value) is mapped to a hypervector with $D$ dimensions called $ID$.
We quantize the intensity value for each peak to Q levels and assign each a base hypervector denoted as $l_j$.
To generate the Q-level base hypervector, we first create a random binary base-hypervector $l_0$ with $D$ dimensions, where $l_0\in \{-1,1\}^D$. The remaining base hypervectors $l_j$  are generated by flipping $D/(2Q)$ bits of the preceding hypervector $l_{j-1}$. This approach ensures that neighboring $l_j$ and $l_{j+1}$ pairs maintain more similarity than pairs that are far apart. Previous studies have shown that the number of quantized levels (commonly selected in the range of Q=16$\sim$32) does not significantly impact the results for this application \cite{HyperOMS}.
Given two sets of hypervectors, a spectrum vector is encoded into a hypervector $h$ (see Figure \ref{encode}) using the following equation.
\vspace{-8pt}
\begin{equation}
    h = Sign\left(\sum_{i\in S} ID_i \otimes LV_i\right)
    \label{encode_equation}
    \vspace{-2pt}
\end{equation}

For each peak in spectrum $S$, we perform element-wise multiplication between the position hypervector $ID_i$ and its corresponding level hypervector $LV_i$ ($LV_i \in \{l_0, \cdots, l_{Q-1}\}$). We then sum the results and apply quantization using the $Sign()$ function to obtain the final binary hypervector $h$.

\subsection{Hamming Similarity Search}
After encoding, we find the most similar reference hypervector to the query hypervector by calculating their similarity. As all hypervectors are binary, HD replaces the cosine similarity with a simpler Hamming similarity. The Hamming similarity, calculated as the number of equal components in vector pairings, is measured by dot product.

\subsection{FDR Filter}
The final step includes applying a false discovery rate (FDR) filter, a widely used method for MS analysis.
It introduces non-existing decoy spectra into the spectral library. The filter then sifts through decoy spectra selected by the search tool. The performance of different search tools can be compared under a fixed FDR threshold.

\section{Hardware Acceleration}

While data preprocessing is typically done offline, encoding and search dominate the algorithm's runtime (over 90\%) and they are memory-intensive tasks that best done in memory.

In the data flow, encoding is the first step conducted in memory, followed by storing encoded hypervectors in memory. Subsequently, a similarity search is performed. However, in this section, we introduce our optimization methods in the reverse order, focusing on the acceleration of the search before delving into the encoding. This decision is due to the encoding's dependence on knowledge introduced during the search acceleration.

\subsection{In-memory Hamming Similarity Search}
The basic operation for hamming similarity search is the MVM between the query hypervector and the reference hypervectors, which can be efficiently accelerated in memory. During the search process, we store each reference hypervector (weight) vertically. In each cycle, a query vector (input) is fed into the array as analog voltages (see Figure \ref{Differential}).

\subsubsection{Weight Mapping}

We use a differential weight mapping scheme, where two cells in a pair together store one number. Compared to the non-differential version, it offers a better solution to challenges arising from non-linearities, such as residual current, RRAM resistance mismatches, and the on-resistance of peripheral switches.

The stored weights in RRAM are encoded reference hypervectors.
A differential pair, comprising two cells in adjacent rows within the same column, together stores one dimension value $W$. Their conductance $g_i^+$ and $g_i^-$ are programmed to the opposite value with respect to the middle level $g_{max}/2$ as follows.
\begin{align}
    \vspace{-2pt}
    g_i^+ = \frac{1}{2}(1+W/W_{max})g_{max} \\
    g_i^- = \frac{1}{2}(1-W/W_{max})g_{max} 
    \label{equation_map}
    \vspace{-2pt}
\end{align}

\subsubsection{Sensing Scheme}
Conventional current sensing \cite{chen201865nm, xue201924} suffers from limited throughput due to constraints in concurrent row driving and increased energy consumption resulting from static current during sensing.  
To mitigate these effects, we apply open-circuit voltage sensing \cite{wan2022compute} with a differential scheme.
\begin{figure}[h]
    \vspace{-10pt}
    \centering
    \begin{subfigure}[t]{0.7\columnwidth}
        \centering
        \includegraphics[width=\textwidth]{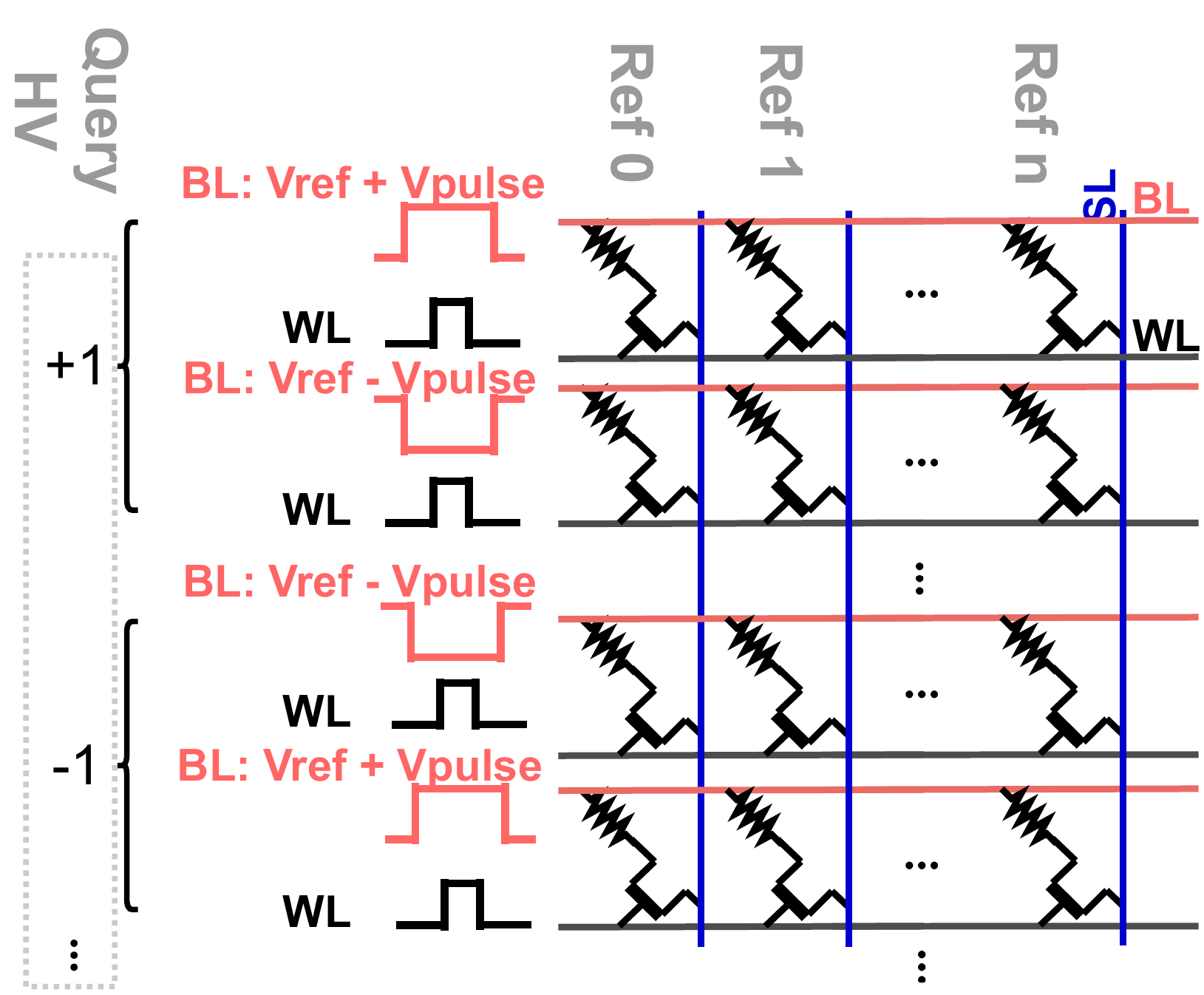}
        \subcaption{In-memory Search}
        \label{Differential}
    \end{subfigure}
    \quad
    \centering
    \begin{subfigure}[t]{0.19\columnwidth}
        \centering
        \includegraphics[width=\textwidth]{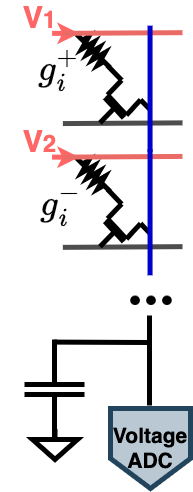}
        \subcaption{Sensing}
        \label{Sensing}
    \end{subfigure}
    \caption{RRAM for MVM}
    \label{rram_search}
    \vspace{-4pt}
\end{figure}

\begin{figure*}[hbt!]
    \vspace{-10pt}
    \centering
    \hfill
    \begin{subfigure}[t]{0.66\columnwidth}
        \centering
        \includegraphics[width=\textwidth]{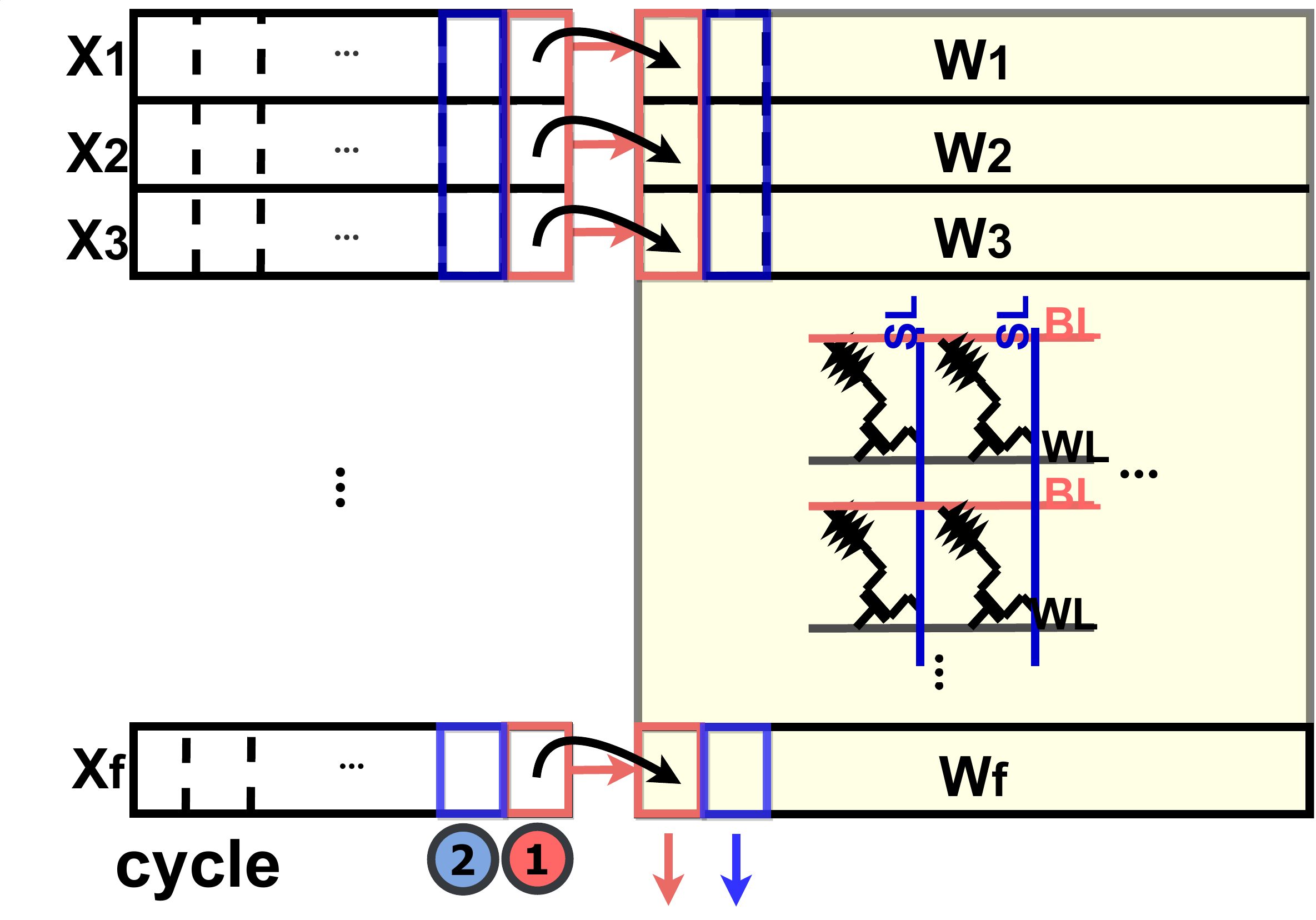}
        \caption{Element-wise MAC in RRAM}
        \label{encoding_elem}
    \end{subfigure}
    \hfill
       \begin{subfigure}[t]{0.66\columnwidth}
        \centering
        \includegraphics[width=\textwidth]{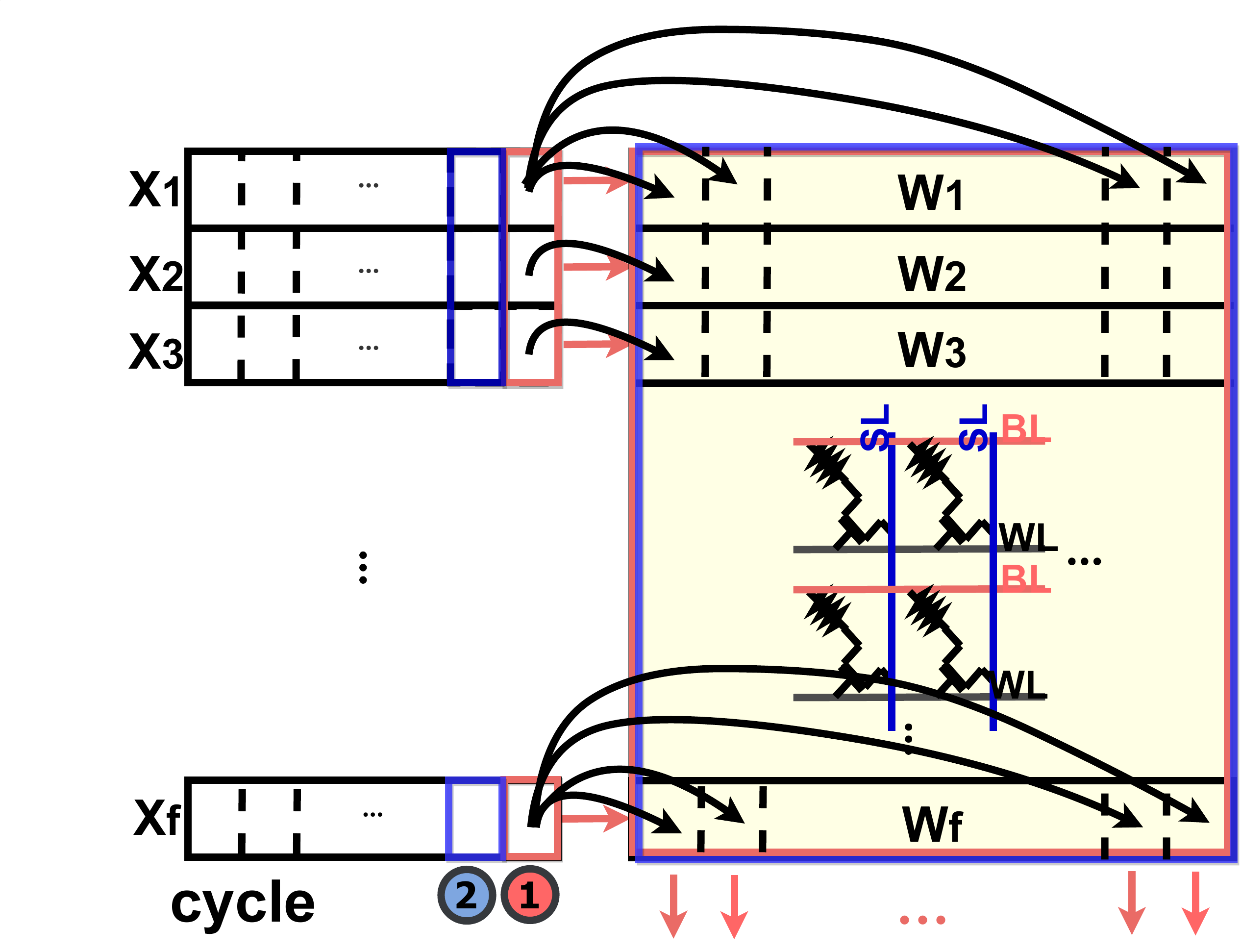}
        \subcaption{MVM in RRAM}
        \label{encoding_mvm}
    \end{subfigure}
    \hfill
    \begin{subfigure}[t]{0.66\columnwidth}
        \centering
        \includegraphics[width=\textwidth]{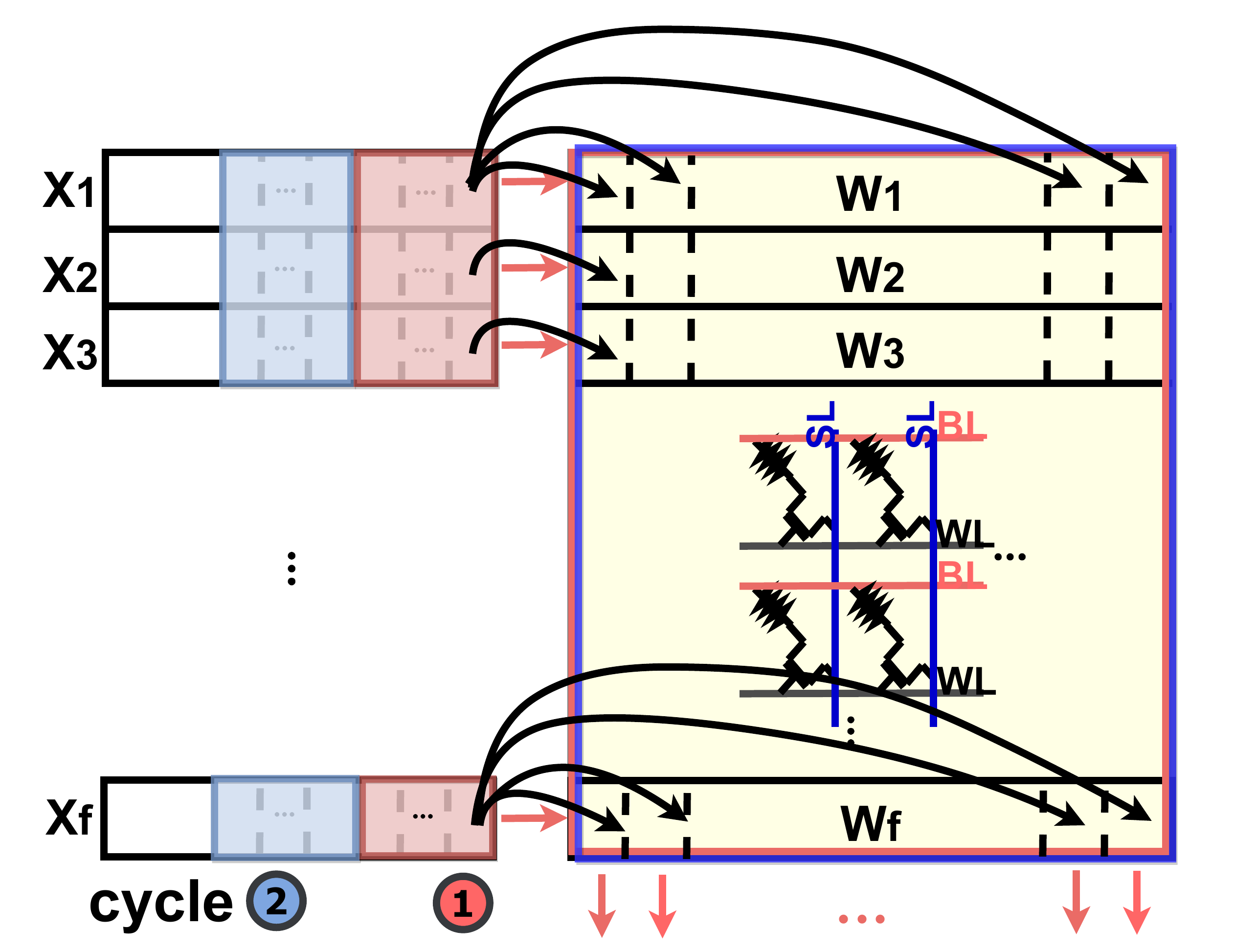}
        \caption{Proposed encoding in RRAM}
        \label{encoding_proposed}
    \end{subfigure} 
    \caption{Encoding in RRAM}
    \label{encoding_all}
    \vspace{-10pt}
\end{figure*}

During MVM, the input (query) hypervector is simultaneously transmitted through differential BL voltages to represent signed inputs (Figure \ref{Differential}). All activated rows contributing to the MAC output, numbering $N$, receive a high signal from the WL. The resulting currents are collected by the capacitor, generating a voltage on the SL according to the following equation.
\begin{equation}
\begin{aligned}
\vspace{-2pt}
C_{} \frac{dV_{SL}}{dt} = \sum_{i=0}^{N}(V_{ref} + V_{pulse}\cdot X_i - V_{SL})\cdot g_i^+ +  \\ 
 \sum_{i=0}^{N}(V_{ref} - V_{pulse}\cdot X_i - V_{SL})\cdot g_i^- 
\end{aligned}
\end{equation}

Once the SL voltage reaches the steady state,
\begin{equation}
\vspace{-2pt}
    V_{SL} = V_{ref} + \frac{\sum_{i=0}^N X_i \cdot (g_i^+ - g_i^-) }{N g_{max}}\cdot V_{pulse}
    \label{SL_steady}
\end{equation}
the ADC then converts the voltage into a digital output. According to Equation \ref{SL_steady}, the resulting output voltage demonstrates a linear relation with the expected MAC output.

\subsubsection{Errors from Memory}

HD exhibits robustness to errors as a result of the the distributed and associative nature of hypervectors. Matched patterns have a significantly higher similarity than unmatched pattern pairs. This substantial difference in similarity acts as a form of tolerance to memory errors.

\subsection{Encoding In Memory}

Encoding maps input data into hypervectors.
The key operation in encoding (see Equation \ref{encode_equation}) is a MAC, which can be accelerated in memory. However, encoding involves element-wise operations, which are challenging to accelerate as effectively as MVM in memory. Based on these observations, optimizing the encoding processes within the memory domain is important.

We use the same weight mapping as in search to convert hypervectors to RRAM conductance, as well as the same differential sensing scheme. RRAM can conduct MAC operations, which involve multiplication of the input and weight on the same row and addition vertically inside a column.  Due to constraints on the orientation of these operations, each position hypervector $ID$ (weight $W$) is stored horizontally, while multiple level hypervectors $LV$ (inputs $X$) are simultaneously fed into the array bit by bit.

\subsubsection{Efficient Encoding}

However, the memory array is less proficient in element-wise operations as compared to MVM.  In MVM, during each cycle when inputs are applied, multiple columns are activated, and each of them generates one MAC output, resulting in multiple MACs per cycle (see Figure \ref{Differential} and \ref{encoding_mvm}). On the other hand, in element-wise operations, only one corresponding output from the array is valid in each cycle (see Figure \ref{encoding_elem}). Consequently, we need more cycles to finish the same amount of MAC computation. Another challenge arises from the fact that, even though only one column of output is valid, it is not possible to selectively activate cells in that specific column. In the memory array, BL and WL are shared by all cells on the same row. When input comes, all cells on that row are driven, consuming power, even though some of them are generating unnecessary results. This makes it more power-hungry compared to MVM.

Based on this observation, we aim to transform the element-wise operation into an MVM-style to enhance throughput and energy efficiency. This is achieved by modifying the way we generate inputs (level hypervectors). Given that the number of base $LV$ hypervectors (Q=16$\sim$32) is much less than the hyperspace size, the choice of $LV$ hypervectors has minimal impact on final results. Instead of randomly generating a base $LV$ hypervector, the $D$-bit hypervector is divided into several chunks, where all bit values within each chunk are identical. Instead of feeding the entire $D$-bit $LV$ hypervectors into the array bit by bit, we now do it chunk by chunk since all values inside one chunk are the same. The number of chunks should be selected based on algorithm-related factors such as HD dimension, the number of base $LV$ hypervectors, as well as hardware-related factors including array size, 
and the column-sharing arrangement for ADCs. Overall, this modification allows all element-wise MAC outputs within one chunk to be obtained in a single cycle, resembling the MVM fashion.

\subsubsection{Multi-bit Hypervector}

In prior HD research \cite{xu2023fsl, imani2019bric, HyperOMS, salamat2020accelerating}, input data is represented by binary hypervectors for further use. However, inspired by the fact that MLC hardware has the capacity to store multi-bits per cell, and considering that synapses in the brain have 4.7-bit precision \cite{bartol2015nanoconnectomic}, there arises a potential that using a multi-bit hypervector scheme could produce better performance.

Instead of generating a binary hypervector $ID \in \{-1,1\}^D$, we apply a multi-bit approach. Each dimension of the $ID$ hypervector could be up to 3 bits, for example, $ID\in\{ -4,-3,-2,-1,1,2,3,4\}^D$. Additionally, this adaptation introduces no additional hardware cost, while raising the bit precision in input level hypervectors $LV$  would increase the overall number of cycles for processing, since inputs are fed into the array in a bit-serial fashion.

\subsubsection{Errors from Memory}
During encoding, final outputs are quantized to binary using the $Sign()$ function, requiring only a low-resolution MAC output. This tolerates errors from memory cells, for example, a single bit flipping would not significantly affect the output. It also reduces ADC design requirements, minimizing computing errors introduced by the ADC.

\subsection{Hypervector Storage}
\label{mapping_storage}
After encoding, we stored the encoded hypervectors in memory. In MLC RRAM, each cell can exhibit $2^n$ levels of conductance, allowing it to store $n$-bit data per cell ($n=1,2,3$). Reference hypervectors used for computation in later stages (Hamming search) are stored in a differential manner, as shown in the previous section. However, to maximize storage capacity, we store all query hypervectors using the following non-differential method.
 We reshape the $D$-bit hypervector into segments of $n$-bits each, forming a new $D/n$-bit vector denoted as $h'$, and map them to unsigned integer values accordingly. These $h'$ are then further mapped onto the RRAM conductances $g$. The following example illustrates how to store a hypervector within 4-level-cell RRAM ($n$=2 bits per cell).
 
\begin{table}[h]
\vspace{-4pt}
  \label{}
  \begin{tabular}{ccc}
     Binary value to store $h\in$  $\Leftrightarrow$ &  Integer $h'\in$ $\Leftrightarrow$ & Conductance $g$ \\
     $\{(-1,-1),\cdots,(1,1)\}^{D/2}$ &$\{0,\cdots,  3\}^{D/2}$ &$g = h'/h'_{max}*g_{max}$\\
\end{tabular}
\vspace{-8pt}
\end{table}

\section{Evaluation}
\subsection{Experiment Setup}

\subsubsection{MLC RRAM Chip Measurement}

\begin{figure}[h]
\vspace{-8pt}
	\centering
 	\begin{minipage}[b]{.49\columnwidth}
		\centering
		\includegraphics[width=\textwidth]{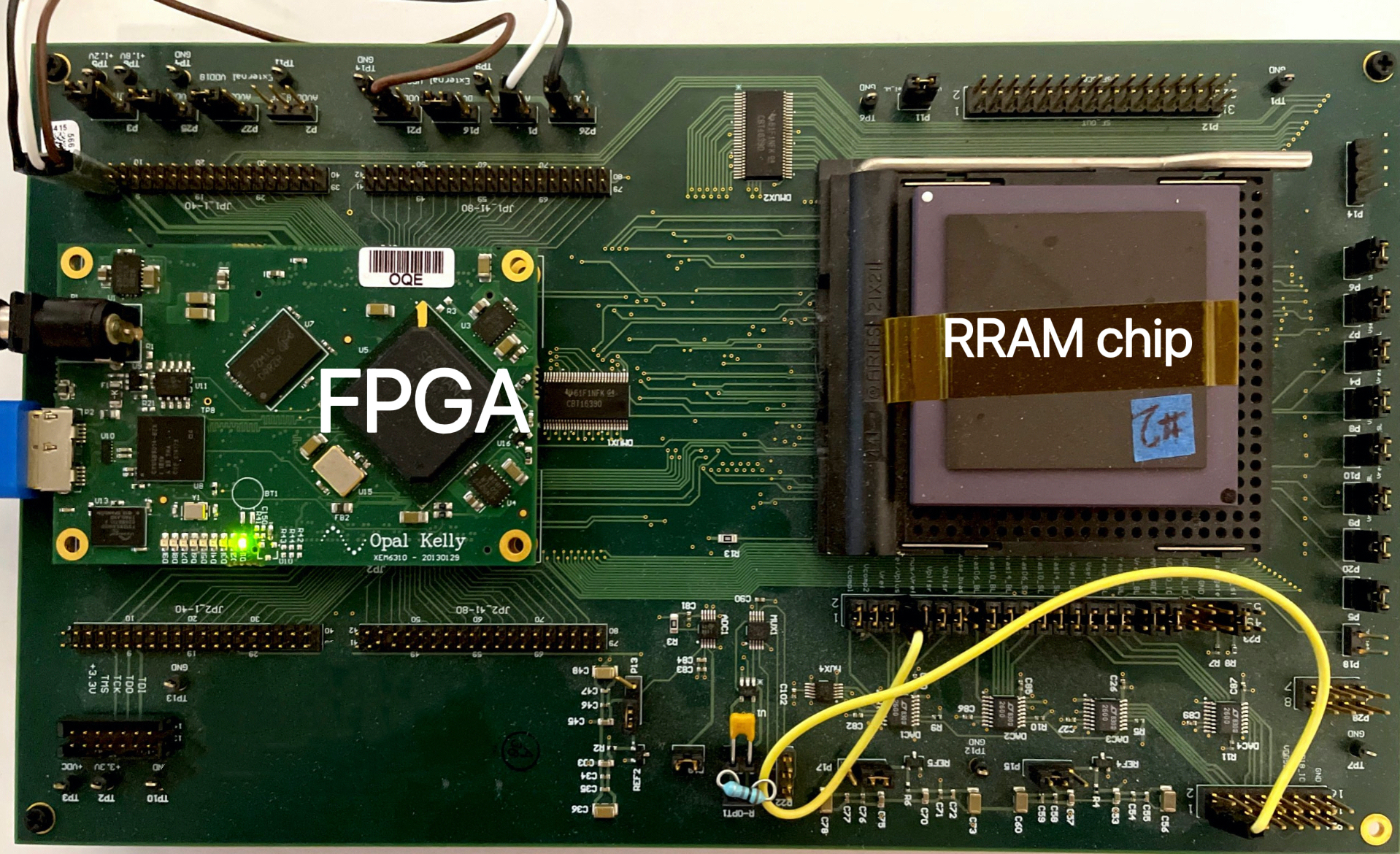}
		\caption{MLC RRAM chip setup}
        \label{chip_setup}
	\end{minipage}
    \hfill
	\begin{minipage}[b]{.49\columnwidth}
		\centering
		\includegraphics[width=\textwidth]{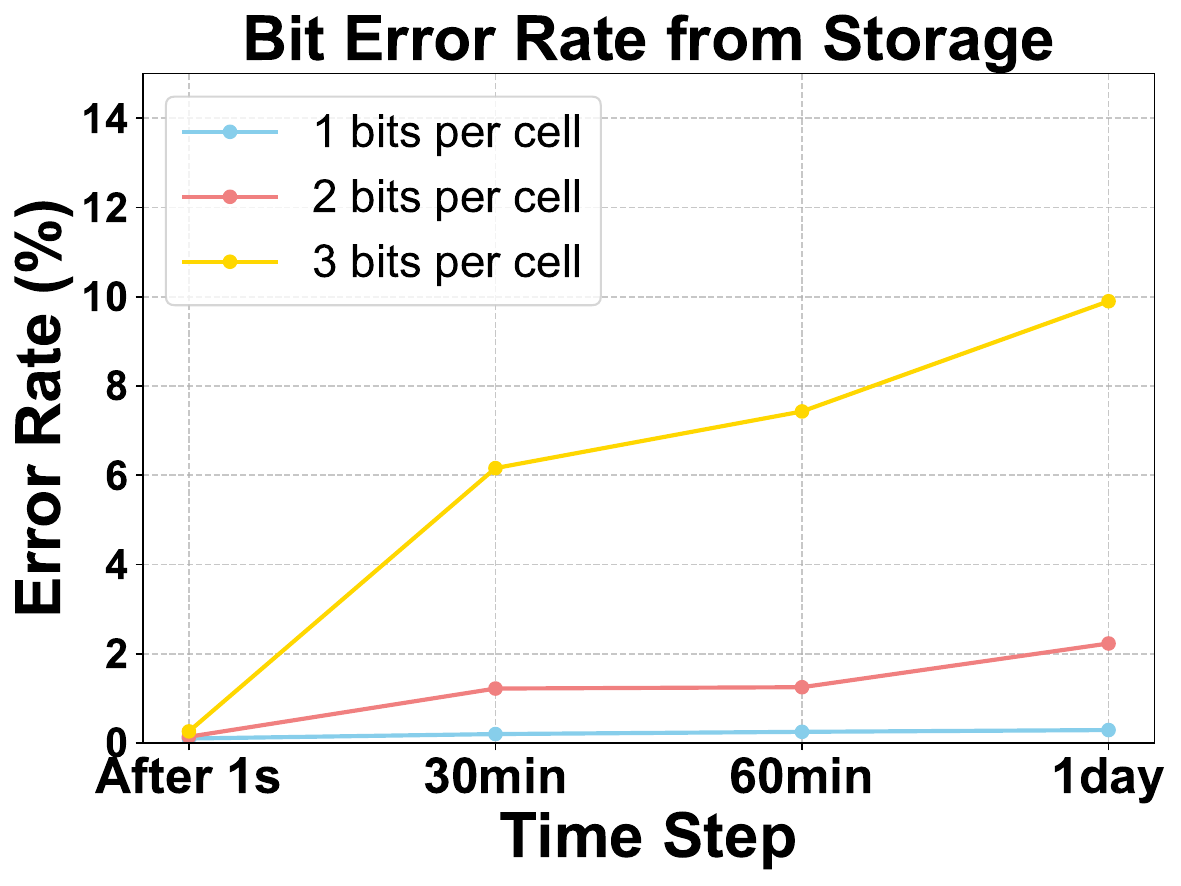}
		\caption{Storage errors}
        \label{rram_store_result}
	\end{minipage}
\vspace{-8pt}
\end{figure}

We tested our algorithm on a fabricated MLC RRAM chip \cite{wan2022compute} in 130nm technology with a total of 3 million RRAM cells. The Xilinx FPGA integrated on an Opal Kelly XEM6310 module serves as the communication bridge between host computer and the chip (Figure \ref{chip_setup}).
The chip loads/writes the input/output hypervectors from/to off-chip text files.

\subsubsection{OMS Workload and Benchmark}

We evaluate the design using two real-world datasets. The first dataset uses iPRG2012 (16k  spectra in total) \cite{small_query} as query and  human HCD yeast library (1M spectra in total) \cite{small_ref} as reference. The second dataset uses HEK293 b1906 (47k spectra in total) \cite{hek} as query and  human spectral library (3M spectra in total) \cite{large_ref} as reference. 
We compare our result against two state-of-the-art OMS work, ANN-SoLo \cite{ANN_SoLo'23} on CPU/GPU and HyperOMS \cite{HyperOMS} on GPU.
Baseline benchmarking is conducted on the NVIDIA GeForce RTX 4090 GPU and the Intel Core i7-11700K CPU, respectively. Parameters for data preprocessing and FDR filtering are presented in Table \ref{OMS_setting}.

\begin{table}
\vspace{-8pt}
  \caption{OMS workload settings}
  \label{OMS_setting}
  \begin{tabular}{ccc}
    \hline
     Dataset & iPRG2012& HEK293 \\
    \hline
     number of query spectra& 16k & 47k\\
    number of reference spectra & 1M & 3M\\
  \hline
\end{tabular}
\vspace{-8pt}
\end{table}

\subsection{MLC RRAM Measurement}

\subsubsection{RRAM for hypervector storage}

\begin{figure}[h]
    \centering
    \includegraphics[width=0.49\textwidth]{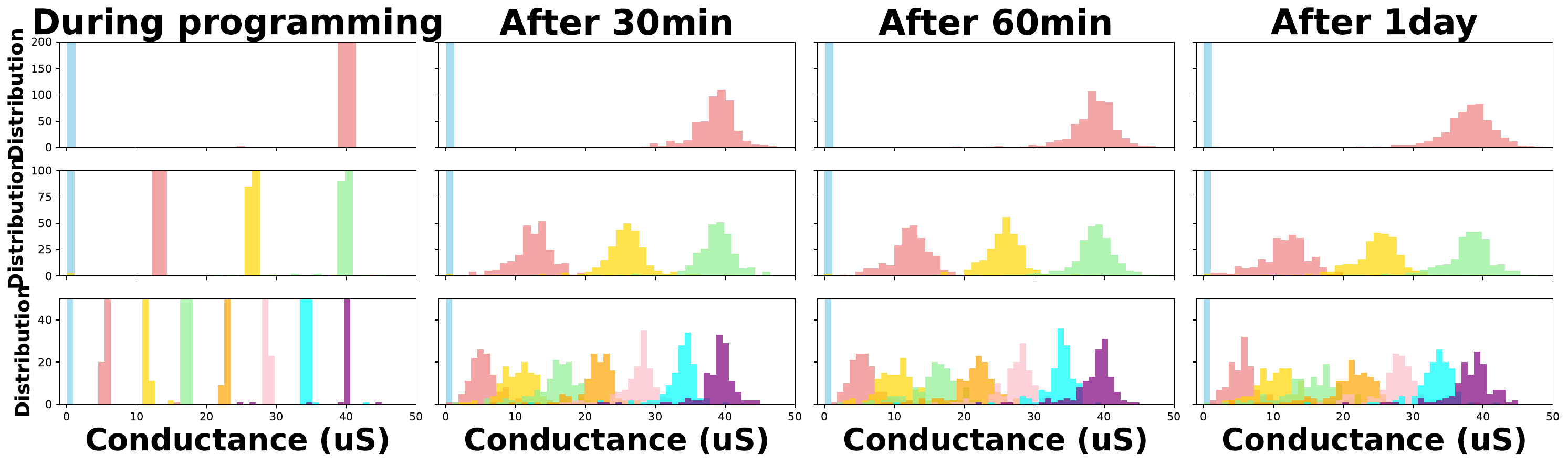}
    \caption{Conductance relaxation effect of 1/4/8-level RRAM}
    \label{MLC_relax}
\end{figure}

RRAM relaxation predominantly occurs at the initial stages, so collecting data after 1 or 2 days does not significantly matter. Therefore, we collect data right after programming, after 30 minutes, 60 minutes, and 1 day, respectively. Figure \ref{MLC_relax} illustrates the histogram of the conductance, from which we compute the bit error rate data in Figure \ref{rram_store_result} based on the mapping method described in Section \ref{mapping_storage}. Our design can store up to 3 bits per cell, leading to a 3x improvement in storage capacity.

In the subsequent sections, all data are collected at least 2 hours after programming to account for RRAM relaxation effects.

\subsubsection{RRAM for computing}

\begin{figure}[h]
\vspace{-8pt}
	\centering
	\begin{subfigure}[t]{0.49\columnwidth}
		\centering
		\includegraphics[width=\textwidth]{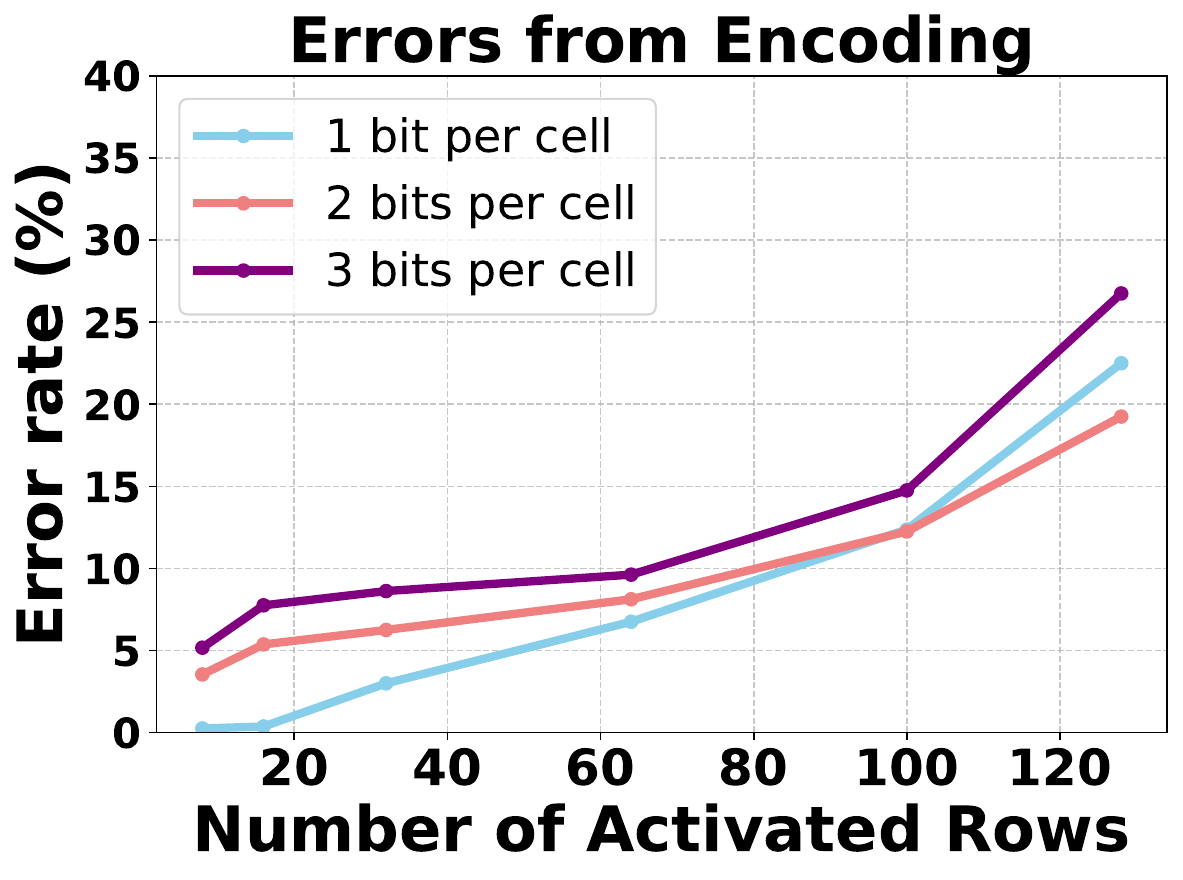}
		\subcaption{Encoding errors}
	\end{subfigure}
    \hfill
	\begin{subfigure}[t]{0.49\columnwidth}
		\centering
		\includegraphics[width=\textwidth]{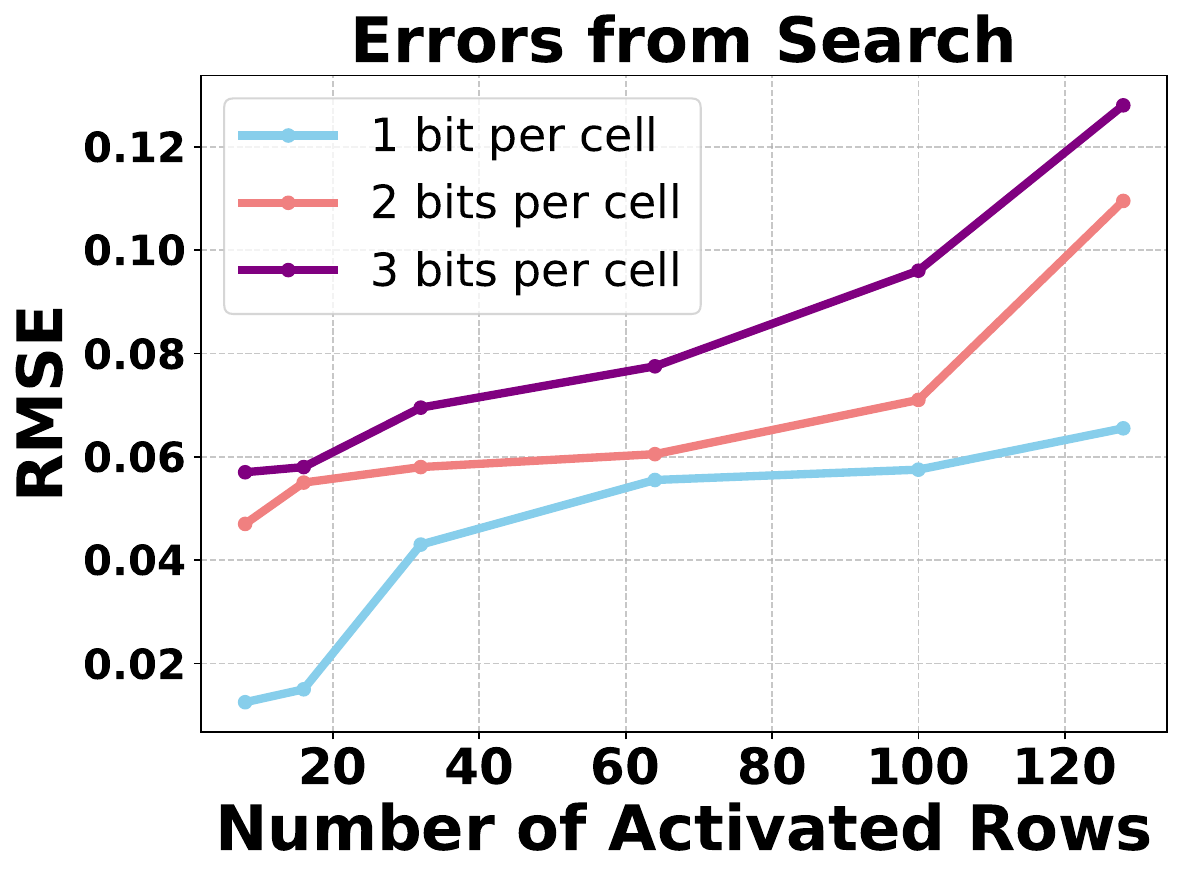}
		\subcaption{Search errors}
        \label{}
	\end{subfigure}
    \caption{Computation errors}
    \label{rram_compute_result}
\end{figure}

For encoding, we compare the binary outputs from RRAM with the corresponding ground truth binary values to calculate bit error rate.
In the case of in-memory hamming search, as the output consists of integer numbers rather than binary, we report the normalized mean square error. Figure \ref{rram_compute_result} illustrates the error rates with 1/2/3 bits storage per cell, corresponding to 2/4/8 level MLC cells, respectively. 
With an increasing number of activated rows, we achieve higher throughput but experience more computation errors.

Compared with the state-of-the-art MLC RRAM design for in-memory computing \cite{li202240}, which can only drive a maximum of 4 rows with 3-level RRAM, our design can activate up to 64 rows (use this setting in following section) with 8-level RRAM, indicating an 16x increase in throughput along with greater storage capacity. The performance gain is due to improvements in RRAM device, design strategy, and benefits from robust HD.

\subsection{OMS results}

\subsubsection{Search quality}

\begin{figure}[h]
\vspace{-8pt}
    \centering
    \begin{subfigure}[t]{0.23\textwidth}
        \centering
        \includegraphics[width=\textwidth]{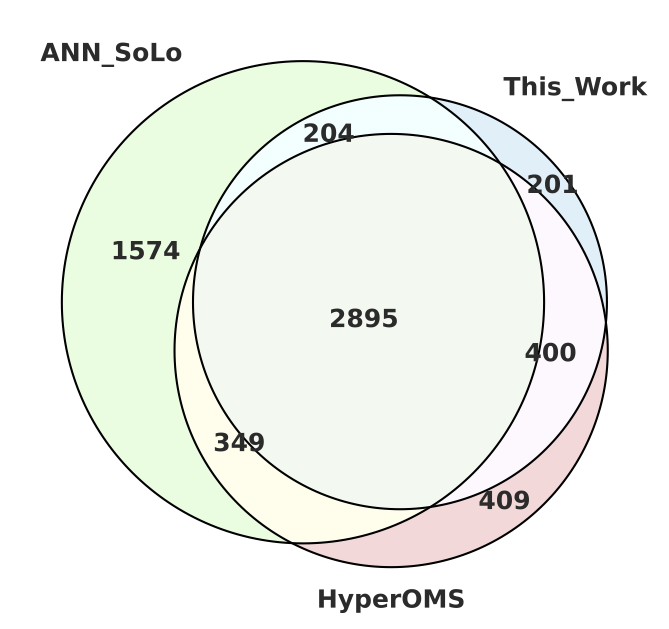}
        \subcaption{iPRG2012 dataset}
    \end{subfigure}
    \begin{subfigure}[t]{0.23\textwidth}
        \centering
        \includegraphics[width=\textwidth]{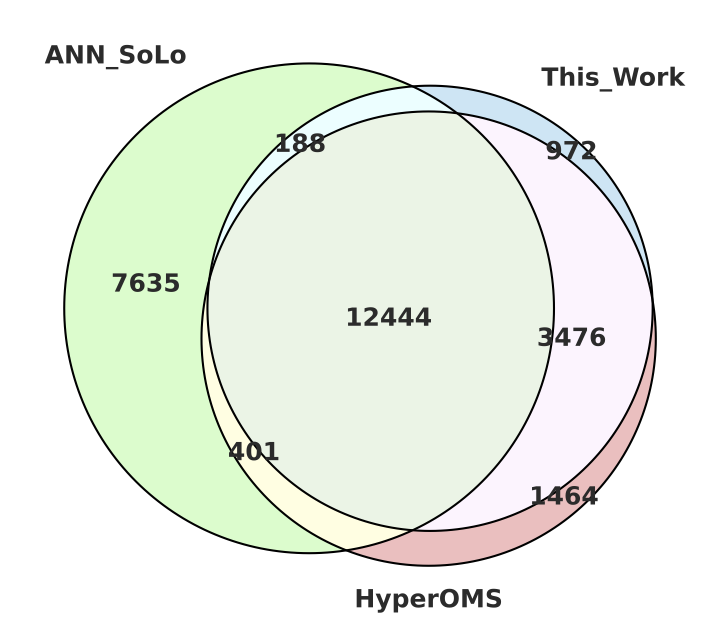}
        \subcaption{HEK293 dataset}
    \end{subfigure}
    \caption{Venn Diagram of identified peptides}
    \label{venn}
\end{figure}

Biological data analysis is complicated, and there is no ground truth data for the search results. Consequently, we compare our results with existing tools.
We set the dimension to be 8k with an $ID$ hypervector precision of 3 bits. Figure \ref{venn} shows the comparisons of the identified peptides. It indicates that the majority of the identified peptides from our work align with those identified by other tools, indicating the validity of our results.

\subsubsection{HD robustness}
\begin{figure}[h]
    \centering
    \includegraphics[width=.5\textwidth]{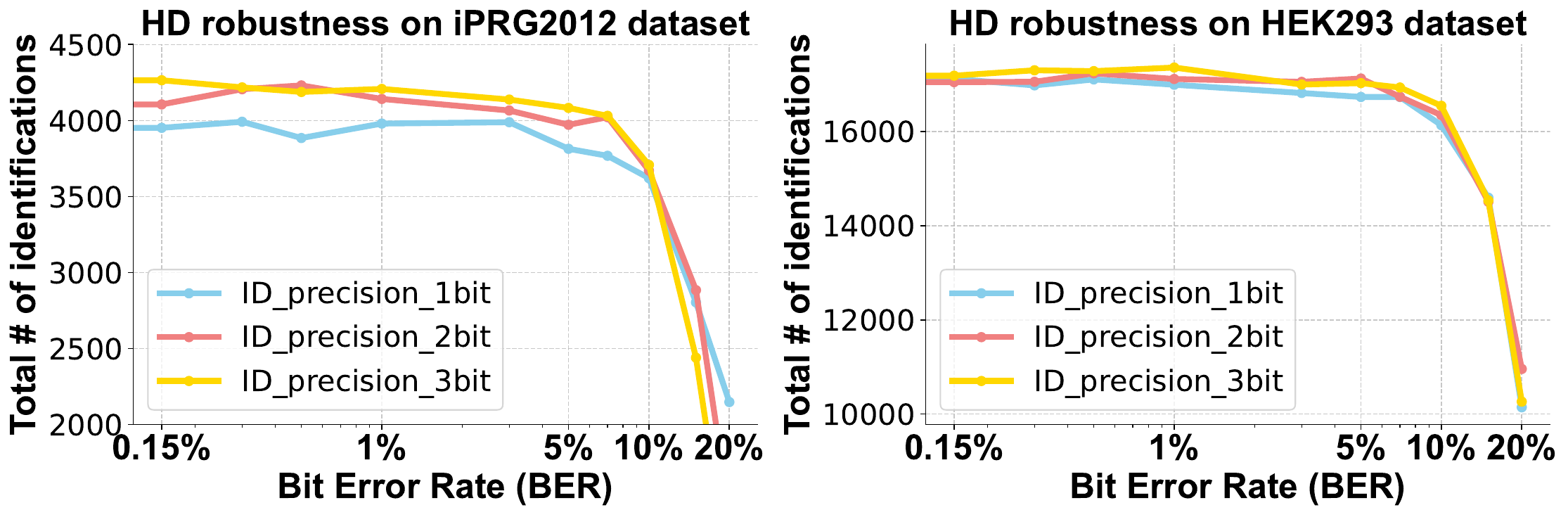}
    \caption{HD robustness}
    \label{BER}
\vspace{-8pt}
\end{figure}

With HD dimension being 8k, we introduce varying levels of bit error rates for encoding and search into the HD algorithm. Figure \ref{BER} shows that our design can tolerate up to 10\% errors. Another notable finding is the enhanced performance achieved through the utilization of multi-bit hypervector scheme.

\subsubsection{Speedup and Energy Improvement}
\begin{figure}[h]
\vspace{-6pt}
	\centering
 	\begin{minipage}[b]{.49\columnwidth}
		\centering
		\includegraphics[width=\textwidth]{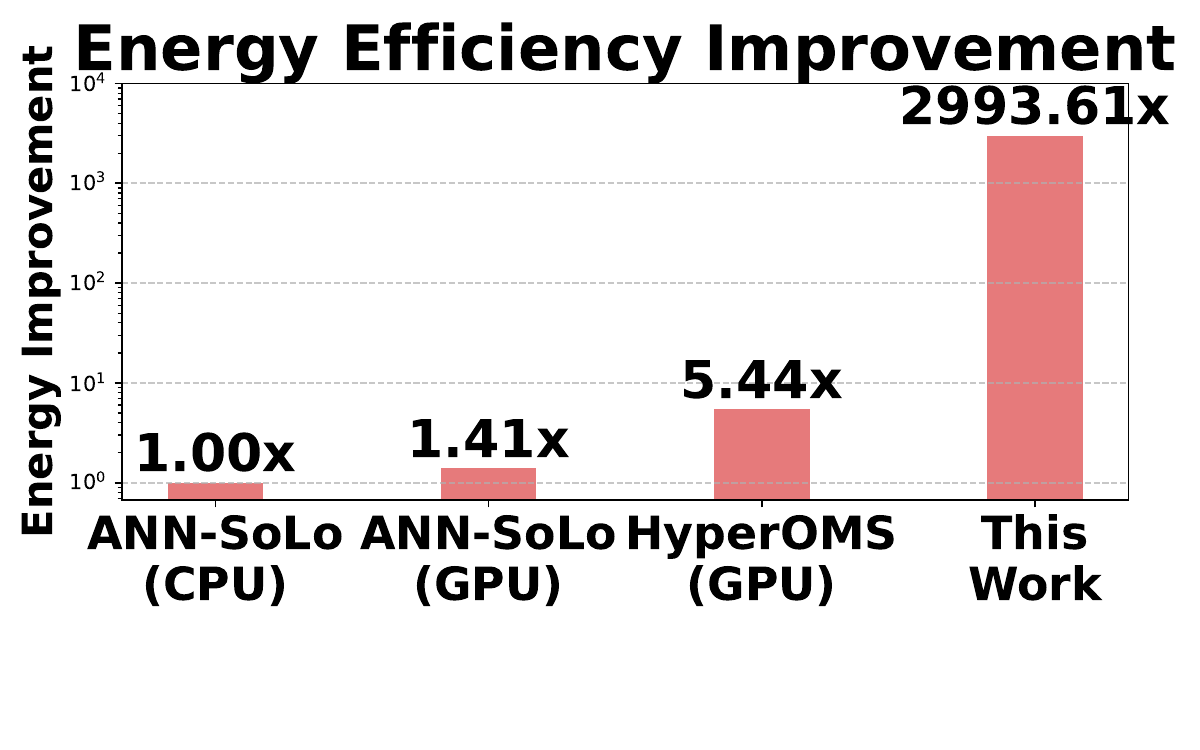}
        \vspace{-20pt}
		\caption{Energy Efficiency}
        \label{energy}
	\end{minipage}
    \vspace{-6pt}
    \hfill
    \vspace{-6pt}
	\begin{minipage}[b]{.49\columnwidth}
		\centering
		\includegraphics[width=\textwidth]{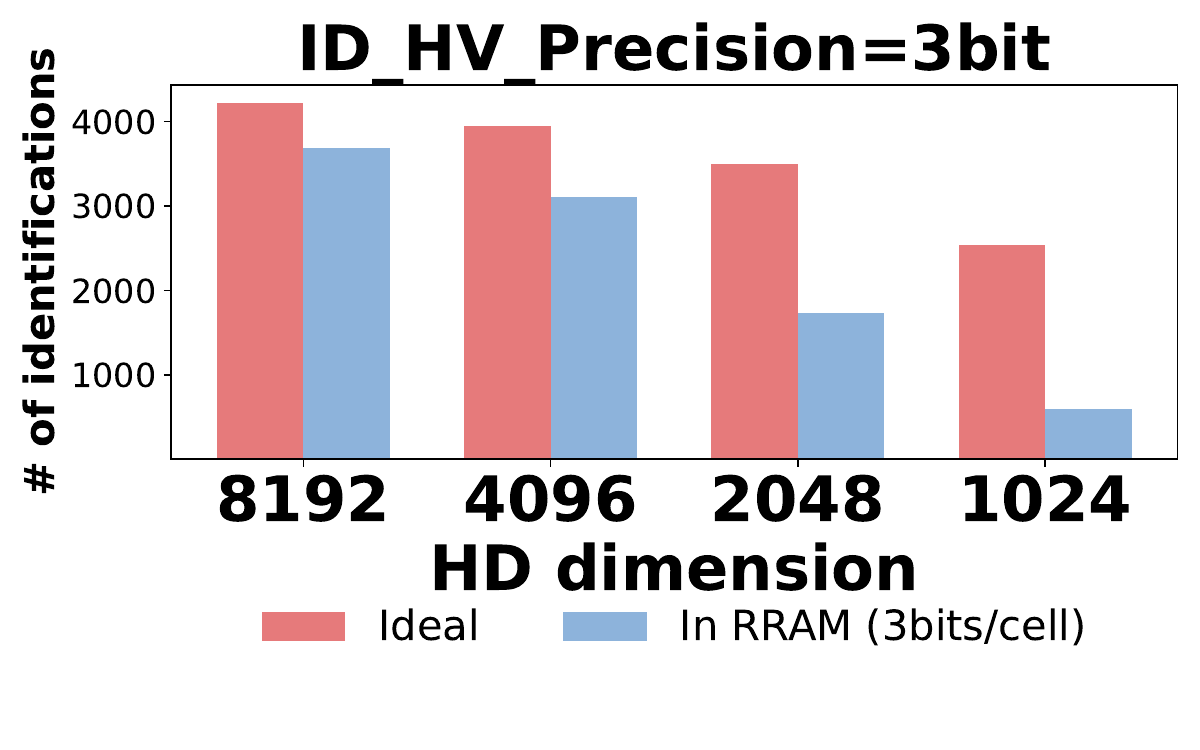}
        \vspace{-20pt}
		\caption{HD Dimensions}
        \label{HD_dimension}
	\end{minipage}
\end{figure}

We simulated the speedup and energy efficiency improvement on iPRG2012 dataset. Our work exhibits 1.7x faster than HyperOMS on GPU, 24.8x/76.7x than ANN-SoLo on GPU/CPU, with 500x-3000x more energy efficiency than state-of-the-art tools (see Figure \ref{energy}). And we expect our performance to scale with more advanced CMOS technology.

\subsubsection{HD dimension}

The HD dimension is a key factor that impacts final results. Lower dimension is more sensitive to noise and exhibits limited separability (see Figure \ref{HD_dimension}). However, an excessively high dimension introduces more computation resource requirements, emphasizing the need for a balanced consideration for target applications.

\section{Conclusion}

In this paper, we propose an accelerator for open modification library searching. We use multi-level-cell RRAM to increase storage capacity by 3x, along with a robust hyperdimensional computing algorithm that can tolerate up to 10\% errors from memory. We accelerate the main stages by computing in memory, leading to 1.7x-76.7x faster processing and 500x-3000x energy efficiency improvement. The functionality of our accelerator has been successfully verified on a fabricated RRAM chip.
Although our accelerator focuses on mass spectrometry applications, the ideas of robust HD and hardware acceleration techniques have the potential for broader applications beyond the realm of mass spectrometry.

\begin{acks}
This work was supported in part by PRISM and CoCoSys, centers in JUMP 2.0, an SRC program sponsored by DARPA, TSMC, and NSF grants 2003279, 1826967, 1911095, 2052809, 2112665, 2112167, and 2100237.
\end{acks}

\bibliographystyle{ACM-Reference-Format}
\bibliography{samples/dac}

\end{document}